\documentclass[12pt]{article}
\usepackage{amsmath}
\usepackage{amssymb}
\usepackage{graphicx,psfrag,epsf}
\usepackage{enumerate}
\usepackage{natbib}

\newcommand{\blind}{1}

\usepackage{url} 
\usepackage{graphicx}
\usepackage[para,online,flushleft]{threeparttable}
\usepackage{float}
\usepackage{xcolor}
\usepackage{siunitx}
\usepackage{enumerate}
\usepackage[font=footnotesize,labelfont=bf]{subcaption}
\usepackage{float}
\usepackage{color}
\usepackage{epstopdf}
\usepackage{amssymb}
\usepackage{listings}
\usepackage{longtable}
\usepackage{authblk}
\usepackage[linesnumbered,ruled,lined,boxed,commentsnumbered]{algorithm2e}
\usepackage{setspace}
\usepackage[normalem]{ulem}
\usepackage[font=footnotesize,labelfont=bf]{caption}
\def\*#1{\mathbf{#1}}
\def\+#1{\boldsymbol#1}

\begin{document}

\def\spacingset#1{\renewcommand{\baselinestretch}%
{#1}\small\normalsize} \spacingset{1}


\if1\blind
{
  \title{\bf MIP-BOOST: Efficient and Effective $L_0$ Feature Selection for Linear Regression}
  \author{\small Ana Kenney\thanks{
    We thank Matthew Reimherr for useful discussions and comments. This work was partially funded by the NIH B2D2K training grant and the Huck Institutes of the Life Sciences of Penn State, and by NSF grant DMS-1407639. Computation was performed on the Institute for CyberScience Advanced CyberInfrastructure (ICS-ACI; Penn State University).}\hspace{.2cm}\\
     Dept.~of Statistics, Penn State University. University Park PA, USA\\
    and \\
    Francesca Chiaromonte \\
    Inst.~of Economics \& EMbeDS, Sant’Anna School of Advanced Studies. Pisa, Italy. \\
    and \\
    Giovanni Felici \\
    Istituto di Analisi dei Sistemi ed Informatica, Consiglio Nazionale delle Ricerche. Rome, Italy.}
    \date{}
  \maketitle
} \fi

\if0\blind
{
  \bigskip
  \bigskip
  \bigskip
  \begin{center}
    {\LARGE\bf {MIP-BOOST: Efficient and Effective $L_0$ Feature Selection for Linear Regression}}
\end{center}
  \medskip
} \fi
\bigskip
\begin{abstract}
\noindent
Recent advances in mathematical programming have made Mixed Integer Optimization a competitive alternative to popular regularization methods for selecting features in regression problems. The approach exhibits unquestionable foundational appeal and versatility, but also poses important challenges. Here we propose MIP-BOOST, a revision of standard Mixed Integer Programming feature selection that
reduces the computational burden of tuning the critical sparsity bound parameter and improves performance in the presence of feature collinearity and of signals that vary in nature and strength. The final outcome is a more efficient and effective $L_0$ Feature Selection method for applications of realistic size and complexity, grounded on rigorous cross-validation tuning and exact optimization of the associated Mixed Integer Program. Computational viability and improved performance in realistic scenarios is achieved through three independent but synergistic proposals. 
\end{abstract}

\vfill

\newpage
\spacingset{1.5} 
\vspace{-0.2in}
\section{Introduction}
\vspace{-0.1in}

Feature Selection methods are a key component of contemporary regression analyses, where they allow researchers to  identifying a subset of relevant or ``active" predictors among an often very large number of available candidates. 

LASSO-type methods (\cite{tibshirani1996regression}, \cite{zou2005regularization}, \cite{meier2008group}) select features solving a convex optimization problem that shrinks the $L_1$ norm of the coefficient estimates vector. Unlike prior best subset and sequential selection approaches (see \cite{hastie2005elements} for an overview), these methods induce sparsity without a direct control on the number of features selected and gained enormous popularity due to the improvements they provided in terms of computational burden, accuracy, or both.
 
More recent proposals adopt Mixed Integer Programming (MIP) to regain direct control on the $L_0$ norm of the coefficient estimates vector, i.e.~the size of the active set, while substantially reducing the computational burden on problems of realistic size through state-of-the-art optimization techniques 
(\cite{bertsimas2016best}, \cite{bertsimas2017logistic}). These proposals engendered a lively debate 
(\cite{hastie2017extended}, 
\cite{mazumder2017subset}, 
\cite{bertsimas2017sparse}), extensive comparisons with the LASSO, and the introduction of ``hybrids" -- e.g., the relaxed LASSO (\cite{hastie2017extended}, see also \cite{meinshausen2007relaxed}) and the penalized MIP (\cite{mazumder2017subset}).

Notwithstanding their reduced computational cost, existing implementations of the MIP approach remain rather burdensome -- and in fact often non-viable for problems comprising thousands or more features, or when performing a rigorous tuning of the sparsity bound (i.e.~of the number of features to be retained). Moreover, whether MIP or LASSO approaches produce better prediction, coefficients estimation and active set identification depends on problem complexity. 
Signal strengths and heterogeneity across features, the degree of true underlying sparsity, and linear associations among features, can all 
have important effects on the relative performance of the two approaches.

According to theory, LASSO can separate active and inactive features only under the irrepresentable condition (\cite{zhao2006model}), which bounds the correlations between them. This may require near-orthogonality between the two sets of features, which is rather unrealistic in many applications. In this regard, \cite{zhao2006model} argued that for universal consistency of the LASSO (i.e.~selection of the true active set) the amount of shrinkage needs to be reduced -- leading to a regularization similar to that of the $L_0$ penalty. To date, the comparative performance of MIP and LASSO approaches in the presence of collinearity and signals that vary in nature and strength is only partially understood. 

The lack of conclusive comparisons may be due to biases arising from computational limitations in tuning. Proper choices of the tuning parameter are critical for the performance of both LASSO and MIP. In LASSO this parameter is a non-negative scalar 
weighing the $L_1$ penalty. In MIP it is the sparsity bound, i.e.~the integer size of the active set represented by the right hand side of a linear integer constraints. This intrinsically harder and more expensive tuning problem is handled with time caps (e.g., the three minute limit in \cite{hastie2017extended}), using a simple training/test regime instead of $v$-fold cross-validation, and/or relying on validation results obtained in simplistic scenarios, where the optimal tuning parameter is easy to ``spot".

Best tuning practices are based on estimating out-of-sample performance, e.g.,~by cross-validation. But for a problem comprising $p$ features, tuning the MIP by standard $v$-fold cross-validation requires solving $v \times p$ instances of the optimization problem (or $v \times n$ when $p>n$) -- with each instance carrying a still substantial computational cost. 

Though theory-based metrics such as AIC or BIC (Akaike and Bayes Information Criteria) can be used to avoid cross-validation, they rely on distributional assumptions and asymptotic properties that may be unrealistic. Interestingly, \cite{miyashiro2015mixed} treat information criteria as the objective in a reformulated MIP which must be solved only once, with the sparsity bound value internally determined -- but this was found to be in fact slower than solving the original MIP over all possible values of the bound. Proposals also exist to improve the computational efficiency of each individual MIP run, usually adopting sub-optimal solutions based on linear relaxations and heuristics  which do not carry a provable quality guarantee (\cite{willis2017l0}, \cite{hazimeh2018fast}). 

We contribute to this debate proposing MIP-BOOST, an enhanced method for $L_0$ Feature Selection based on extensions that significantly improve computational efficiency as well as effectiveness of the MIP approach. These extensions leverage specific characteristics of the class of optimization techniques used for solving $L_0$ Feature Selection problems.

First, we propose a {\it novel bisection procedure} designed specifically for tuning the sparsity bound, which significantly reduces the needed number of evaluations. While this modified bisection can in principle be employed in other optimization settings, we focus on documenting its remarkable impact in cutting the computational burden of the MIP approach.

Second, we propose a {\it novel cross-validation scheme} that exploits the structure of the MIP and of the simplex algorithm used for its solution to significantly reduce the computational effort of repeating calculations across folds. We also employ warm starts and surrogate lower bounds 
within a state-of-the-art MIP solver to further cut running times.

Third, we complement the MIP with {\em whitening} (\cite{kessy2018optimal}). This is a pre-processing step that, by handling feature collinearities, can both reduce computational burden and improve solution quality. Whitening can be applied prior to any feature selection technique but it benefits MIP more than it does other approaches.

We test MIP-BOOST on an extensive body of synthetic data and on a Diabetes data set already analyzed in \cite{efron2004least}, demonstrating a reduction of several orders of magnitude in the computational burden of MIP and a performance on par with or better than that of LASSO-type methods. Our results, in conjunction with other recent literature, establish $L_0$ Feature Selection as a solid and viable alternative for a wide spectrum of large and complex regression problems. 

The remainder of the article is organized as follows:
Section~2 provides technical background, Section~3 introduces our proposals, Section~4 presents our simulation study and the application to Diabetes data, and Section~5 contains final remarks and perspectives. 

\vspace{-0.2in}
\section{Background}
\vspace{-0.1in}

Consider a linear regression of the form $\+Y = \*X\+\beta +\+\epsilon$, where $\+Y \in \mathbb{R}^n$ is the vector of response values, $\*X \in \mathbb{R}^{n \times p}$ the design matrix containing the values of $p$ predictors, $\+\beta \in \mathbb{R}^p$ the vector of regression coefficients, and $\+\epsilon \in \mathbb{R}^n$ the vector of errors. Throughout the article we assume that the data is centered and omit the intercept. 

In the following, we  describe the standard LASSO approach and set the stage for our proposals introducing Mixed Integer Programs (MIP) and their use to formulate the best subset selection problem. As will appear clearly, the key difference between the MIP and LASSO formulations is the choice of penalization: While MIP uses the $L_0$ norm, 
bounding the size of the subset (the number of selected features), LASSO uses the $L_1$ norm, 
bounding the size of the coefficient vector. 

\vspace{-0.2in}
\subsection{Feature Selection with the LASSO}
\vspace{-0.1in}

LASSO (\cite{tibshirani1996regression}) is a regularization method commonly used for feature selection due to its computational speed and ability to induce shrinkage and fit parameters simultaneously. It solves the constrained optimization
\vspace{-0.15in}
\begin{align}\label{LASSO}
\setstretch{0.65}
&\min_{\+\beta}\ \ \frac{1}{n}||\+Y-\*X\+\beta||_2^2\\
& ||\+\beta||_1 \leq \lambda. \nonumber
\end{align}
\vspace{-0.50in}

\noindent 
where the $L_1$ norm is used to bound the size of coefficient vector by the 
scalar $\lambda \geq 0$. LASSO does not explicitly solve best subset selection 
nor directly control sparsity; it leverages a simple optimization problem comprising continuous variables and convex constraints. 

Many fast algorithms have been proposed for solving \eqref{LASSO}; the most commonly used is a variation of coordinate descent from \cite{friedman2010regularization} (a very efficient implementation is provided in the \texttt{glmnet} R-package). Solution quality depends on the choice of the tuning parameter $\lambda$, but computational speed makes this straightforward: one generates a fine grid of $\lambda$ values ranging from maximal to minimal shrinkage (all
coefficients are forced to $0$ on one extreme and take their OLS values on the other), computes a cross-validation error on each grid value, and selects the minimizer. Another choice, which induces more shrinkage, is the largest $\lambda$ with cross-validation error within one standard deviation from the minimum ({\em parsimonious} LASSO; \cite{hastie2005elements}). 

Overall, notwithstanding the efficient tuning made possible by computational speed, LASSO is known for producing solutions with a high number of false positives. In fact, \cite{tibshirani2011regression} recommended it as a \emph{screen} (instead of a selection procedure), because under certain conditions it achieves the sure screening property (\cite{fan2008sure}).
\vspace{-0.2in}
\subsection{Mixed Integer Programs}
\vspace{-0.1in}

In Mixed Integer Programs (MIP), an objective function is optimized under a collection of constraints -- including constraints that restrict some of the variables to integer values. A well studied class of MIPs are Mixed Integer Quadratic Programs (MIQP), where a quadratic objective function in $\*v\in \mathbb{R}^p$ is optimized under linear constraints, lower and upper bounds on the feasible values of the solution, and integrality constraints. In symbols
\vspace{-0.5in}
\begin{align}\label{genMIQP}
\setstretch{0.65}
&\min_{\+v\in \mathbb{R}^p}{\+v^T\*Q\+v + \+q^T\+v}  \\
&\*A\+v \leq\+b \ \ \ ,\ \ \ \+l \leq \+v \leq \+u \nonumber \\ 
&v_j \in \mathcal{S} \subset \mathbb{Z} \text{ for } j \in \mathcal{J} \subset \{1,2,...,p\} \nonumber
\end{align}

\vspace{-0.2in}
\noindent
where $\*q \in \mathbb{R}^n$, $Q \in \mathbb{R}^{n \times p}$ (pos.~semidefinite), $A \in \mathbb{R}^{n \times p}$, $\*b \in \mathbb{R}^n$, $\*l,\*u \in \mathbb{R}^p$, $\mathcal{S}$ and $\mathcal{J}$ are given. 
They extend an even better studied class of MIPs; Mixed Integer Linear Programs (MILP), where also the objective function is linear. In turn, MILPs extend the simplest class of programs; Linear Programs (LP), where integrality constraints are absent. 

While LPs of any size can be solved very quickly, the integrality constraints in MILPs and MIQPs pose serious computational challenges that have been tackled with dramatic improvements in optimization solvers (see \cite{bertsimas2016best} for an overview). At present, the most effective way to solve a MIP is to relax the integrality constraints and then combine implicit enumeration methods, such as Branch \& Bound, with the addition of constraints that approximate integrality in the relaxed problem, such as Branch \& Cut (see \cite{Schrijver1986}) for an introduction). The overarching goal of recent contributions (e.g.,~\cite{bertsimas2016best}) and of our own work is indeed to harness this extremely powerful machinery for the purpose of feature selection.

\vspace{-0.2in}
\subsection{Best Subset Selection as a Mixed Integer Quadratic Program}
\vspace{-0.1in}

For any given integer $k \leq p$, best subset selection minimizes the regression error using at most $k$ features. This problem has the same structure of \eqref{LASSO}, namely 
\vspace{-0.2in}
\begin{align}\label{bestsubset}
\setstretch{0.65}
&\min_{\+\beta}\ \ \frac{1}{n}||\+Y-\*X\+\beta||_2^2\\
&||\+\beta||_0 \leq k \nonumber
\end{align}
\vspace{-0.50in}

\noindent
but it restricts the subset size directly through the $L_0$ constraint $||\+\beta||_0 = \sum_{i=1}^p I(\beta_i \neq 0) \leq k$. \cite{bertsimas2016best} 
adopted a reformulation based on a
MIQP of the form \eqref{genMIQP} introducing a binary vector $\+z \in \{0,1\}^p$ that indicates weather each feature is selected or not ($z_j = 1$ or $0$): 
\vspace{-0.2in}
\begin{align}\label{miqp}
&\min_{\+\beta,\+z}\ \ \frac{1}{n}||\+Y-\*X\+\beta||_2^2 \\
&-\mathcal{M}z_j \leq  \beta_j \leq \mathcal{M}z_j \ \ j=1,\ldots,p
\ \ \ ;\ \ \ 
\sum_{j=1}^p z_j\leq k \nonumber \\
&\beta_j \in \mathbb{R}\ \ , \ \ z_j \in \{0,1\} \ \ j=1,...,p . \nonumber
\end{align}
\vspace{-0.45in}

\noindent
Here the parameters $\mathcal{M}$ and $k$ play very important roles. Constraining each $\beta_j \neq 0$ to be within $[-\mathcal{M}; \mathcal{M}]$ is referred to as the ``bigM'' trick, and the proper choice of $\mathcal{M}$ is an art. 
A sufficiently large $\mathcal{M}$ guarantees that solutions to \eqref{miqp} are also solutions to \eqref{bestsubset}, but an excessively large $\mathcal{M}$ can affect solvability -- mainly for numerical reasons. Using separate bounds $\mathcal{M} = (\mathcal{M}_1,\ldots,\mathcal{M}_p)'$ for each coefficient increases flexibility in the estimation of $\+\beta$.

\cite{bertsimas2016best} provide a few options for selecting $\mathcal{M}$, as well as a clever formulation that does not require its {\em a priori} choice.
In our implementation we set separate and fairly wide bounds utilizing Ordinary Least Squares (OLS) estimates.
The parameter $k$ has a less technical and rather crucial role for guaranteeing accurate feature selection. It bounds the size of the subset in \eqref{miqp}, directly controlling sparsity. If $k$ is too small, we miss important signals; if it is too large, we overfit the regression. For large $p$, choosing $k$ by standard tuning methods such as 10-fold cross-validation across its entire range $\{1,...,p\}$ is computationally very demanding, if not prohibitive -- even with state-of-the-art solvers. Assuming that the regression is very sparse, this issue can be mitigated by restricting the range of $k$ at the outset (e.g.,~exploring only $k \in \{0,...,10\}$ even though $p$  is in the range of thousands).
In many applications though this may not be appropriate. 

\vspace{-0.2in}
\section{Our Proposals}
\vspace{-0.1in}
\label{sec-methods}

Here we present MIP-BOOST, which provides an efficient and effective way to deploy the MIP machinery for $L_0$ feature selection. 
It is based on three extensions to the classic MIP 
that together contribute to large improvements in execution time and solution quality. 

The first is a modified bisection procedure that limits the grid search required for tuning; the second is an integrated scheme that reduces the computational burden of individual MIP solutions in a $v$-fold cross-validation; and the third is a whitening pre-processing step that handles feature collinearities. Each targets a different aspect of the computational demands of feature selection.

\vspace{-0.1in}
\subsection{Bisection with Feelers}
\label{BWF}
\vspace{-0.1in}

\begin{figure}[H]
    \centering
        \begin{subfigure}[t]{.49\textwidth}             \includegraphics[scale=0.25]{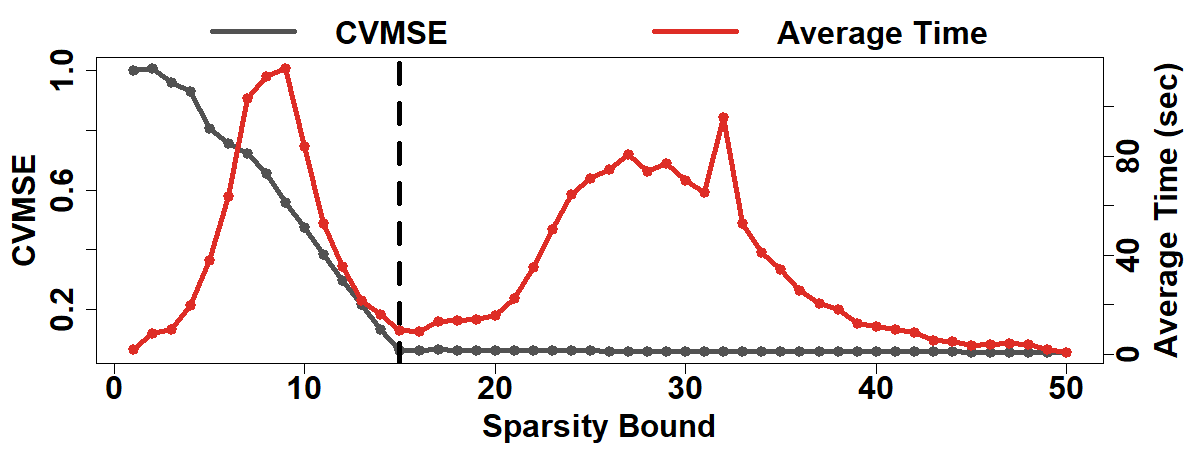}
              \caption{}\label{cvmse_full}
          \end{subfigure}%
          ~
          \begin{subfigure}[t]{.49\textwidth}
       \includegraphics[scale=0.25]{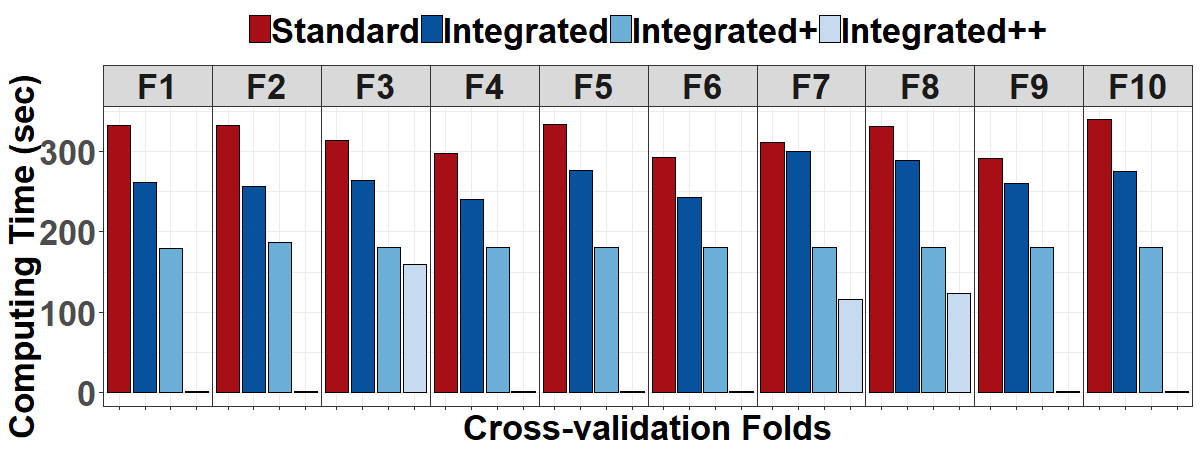}
              \caption{}\label{inncv}
          \end{subfigure}
\caption{\textbf{(a)} Cross-validation Mean Squared Error (CVMSE) and average computing time (over the $10$ folds) across all values of $k$, for a simulation scenario with $n=500$ observations, $p=50$ features, and $k_0=15$. The dashed line marks the true bound. \textbf{(b)} Bar chart of computing times across the $10$ folds when applying integrated, integrated+, integrated++, and standard cross-validation (see Section~3.2) at $k=5$ for a simulation scenario with $n=500$ observations, $p=1000$ features, and $k_0=10$.
}\label{cvmseInncv}
\end{figure}

Let $k_0$ be the true and unknown number of active features. Fixing an appropriate bound, i.e.~a good estimate of $k_0$ based on the data, is critical for the effectiveness of feature selection. This can be done by evaluating out-of-sample regression performance via $v$-fold cross-validation, and comparing it across all $k$ values from $1$ to $p$ (or $n$ when $p>n$). For example, the black curve in Figure~\ref{cvmseInncv}(a) shows 10-fold cross-validation Mean Squared Errors (CVMSE) obtained solving MIPs at $k=1,...,p$ on simulated data with $n=100$, $p=50$ and $k_0=15$ (see Section~4 for details). Ideally, the CVMSE curve would decrease (as one adds more active features) and then steadily increase (as one overfits adding inactive features), identifying $\hat{k}_0$ as a {\em sweet spot}.
But in many practical applications, especially when features are highly correlated and/or signals are weak, the CVMSE curve can have an {\em elbow} shape as in Figure~\ref{cvmseInncv}(a) (or yet more complicated, fluctuating behaviors as in  Figure~3 of the Supplement). Note that non-ideal behaviors of the CVMSE curve can be starker when plotting against discrete sparsity bound values, as opposed to plotting against a fine grid of $\lambda$ values when tuning the LASSO. Our objective is thus to reliably identify a global minimum (a sweet spot) or the smallest, most parsimonious among a ``stretch" of approximately equivalent bound values (an elbow) in a non-ideal CVMSE curve. And the problem is the computational burden of cross-validating across the whole range of $k$: if $p$ is large, this may be unviable because we need to solve as many as $v \times p$ MIPs (or $v \times n$ for $p>n$). For very sparse regressions, say $k_0 \leq 10$, it may still be feasible to solve $v$ MIPs for $k=1,2 \ldots$ until it becomes clear that we are past a sweet spot or an elbow in the CVMSE curve. However, if the regression is not very sparse, or if it is only {\em proportionally} sparse but $p$ is very large, this is still unviable; e.g.,~if $p=5000$ and $k_0=10\% \times p=500$, we need to cross-validate more than $500$ values of $k$ (solve more than $v \times 500$ MIPs) to identify a good estimate of the bound.
\vspace{-.5cm}


If we could count on the CVMSE curve having a single global minimum and a clear quasi-convex shape, we could reduce computational burden with standard 1D search methods that do not require first or second order information. These search the whole range $\{1,\ldots, p\}$ (or $\{1,\ldots, n\}$ if $p>n$) of $k$ but evaluate only a small portion of its values. However, they can be rather ineffective for identifying the elbow of a curve shaped as in Figure~\ref{cvmseInncv}(a) since gains in solution quality, no matter how negligible, push the search towards adding inactive variables. Thus, we propose a procedure that modifies a standard bisection search (\cite{cormen2009introduction}) as to avoid chasing minor gains. We call this \textbf{Bisection with Feelers (BF)}. BF can effectively hone into an elbow, as well as a sweet spot, and it does so efficiently; as a 
bisection, it limits the number of MIPs to be solved to around $v \times \log_2(p)$ (or $v \times \log_2(n)$ when $p>n$).

We begin by initializing $a=a_0$ and $c=c_0$, the two end points of the search interval and splitting at the midpoint $b$ to get $[a,b]$ and $[b,c]$. We can naively set $a_0=1$ and $c_0=p$ (or $c_0=n$ when $p>n$). A better option is to utilize an inexpensive method to set a more informed upper bound. For our experiments in Section~4 we used the LASSO which is known to provide a good upper bound, retaining true positives (albeit often along with false ones; see Section~2.3). Our criterion to decide which of the two intervals to search next is a standardized marginal change in cross-validation error (a slope). Let $M_k$ be the set of features selected by the MIP with a generic $k$, and $f(k)$ the corresponding cross-validation error; the improvement that occurs between, say, $x$ and $y$ is $\Delta f(x,y) = - \frac{1}{f(x)}\frac{f(y)-f(x)}{(y-x)}$, which represents the percentage change per feature added in $[x,y]$ (the negative sign makes positive and higher values of $\Delta f(x,y)$ preferable). We solve for $M_{a}$, $M_{b}$, and $M_{c}$ and evaluate the criterion for $[a,b]$ and $[b,c]$. With a given improvement threshold $\delta>0$, $\Delta f(b,c) \leq \delta$ suggests that $b$ is already past the elbow of the curve, so we search $[a,b]$. Conversely, if $\Delta f(b,c) > \delta$, we search $[b,c]$. To fix a reasonable $\delta$ it helps to think of it as a ``baseline percent improvement" to the starting point of each interval; in our experiments in Section~4 we use 1 or 5\% -- meaning a gain of less than 1 or 5\% is considered negligible. 

In Section~6 of the Supplement we 
provide a proof of convergence for this procedure under some regularity assumptions. Notably, the proof covers the case of a CVMSE curve with multiple elbows and converges to the one with the lowest CVMSE. However, especially under particularly noisy scenarios, 
the CVMSE curve may show fluctuations that break the elbow(s) trend. Thus, we favor sparsity and if $\Delta f(a,b) \leq -\delta$, so $f(b)$ is sufficiently higher than $f(a)$, we search $[a,b]$ regardless of the value of $\Delta f(b,c)$. This indicates that the curve actually re-increases past a minimum occurring before $b$, so that candidate values in $[b,c]$ are subject to overfitting.

The process continues until the bisection converges and outputs a tentative estimate $\hat{k}_{0}$. This though is not necessarily our end result: again to help tackle irregularities in the 
CVMSE curve, we next send ``feelers" around $\hat{k}_{0}$, computing $\Delta f(\hat{k}_{0}-l_f,\hat{k}_{0})$ and $\Delta f(\hat{k}_{0},\hat{k}_{0}+l_f)$, where $l_f$ is the feelers' ``radius" (in our experiments in Section~4 we fix $l_f=1$ and check only immediately to the left and right of $\hat{k}_o$). 
If both are $\leq \delta$, suggesting that $\hat{k}_{0}$ does not provide a sufficient improvement relative to its surroundings, we restart the search setting $a=1$ and $c=\hat{k}_{0}-l_f$. But if $\Delta f(\hat{k}_{0},\hat{k}_{0}+l_f) > \delta$, suggesting that improvement is possible to the right of $\hat{k}_{0}$, we restart the search setting $a=\hat{k}_{0}+l_f$ and leaving $c$ to its initial value.
 
Due to the feelers portion, the number of evaluations in BF may exceed $\log_2(c_0)$, depending on how many restarts we allow and how far past the elbow we are each time. In our experiments we typically restart at most once; nevertheless, we set a bound $itermax$ on the number of iterations to avoid excessive computing (Algorithm 1 in the Supplement provides pseudocode for BF).

\vspace{-.2in}
\subsection{Integrated Cross-Validation}

\vspace{-0.1in}
\noindent
The red curve in Figure~\ref{cvmseInncv}(a) shows average computing times (over cross-validation folds) across $k$ values. Some are much higher than others. As expected, very small or very large $k$'s result in faster convergence rates for the optimization solver (see Section~4.1). Convergence slows down as we move inward and speeds up again around the true $k_0$ -- which is of course unknown in applications. What if, in addition to reducing the number of MIP runs required to tune $k$ (Section~3.1), we could make the most expensive ones cheaper? 

For any given fold, solving the MIP amounts to solving a relaxation of the complete problem where constraints on withheld observations are ignored (i.e.~relaxed). Thus, when the focus shifts from one fold to another, it may be very easy to recover feasibility of the relaxed solution using a dual approach -- a step typically integrated in MIP solvers. We call this {\bf Integrated Cross-Validation} (ICV). For each fold $\mathcal{F}$, we modify \eqref{miqp} relaxing constraints on withheld observations through a parameter $\mathcal{\tilde{M}}$ whose use is similar to that of $\mathcal{M}$ in Section~2. In symbols, we set $m_i = \mathcal{\tilde{M}} \times I(i \in \mathcal{F})$, $i = 1,\ldots,n$ and consider
\vspace{-0.15in}
\begin{align}\label{incv}
&\min_{\+\beta,\+z,\+e} \ \ \sum_{i=1}^n e_i^2 \\
&-e_i-m_i \leq  Y_i -\sum_{j=1}^p \beta_j X_{ij} \leq e_i+m_i 
\ \ i=1,\ldots,n \nonumber \\
&-\mathcal{M}z_j \leq  \beta_j \leq \mathcal{M}z_j 
\ \ j=1,\ldots,p
\ \ \ ,\ \ \ 
\sum_{j=1}^p z_j\leq k \nonumber \\
&\beta_j \in \mathbb{R}\ \ ,\ \ z_j \in \{0,1\} 
\ \ j=1,\ldots,p 
\ \ \ ,\ \ \
m_i = \mathcal{\tilde{M}} \times I( i \in \mathcal{F}) . \nonumber
\end{align}
\vspace{-0.45in}

\noindent
In this formulation subsequent folds are bound to be computationally cheaper than the first, that
serves as a starting point, or warm start, for the others.
However, for very expensive runs even the first fold can be computationally taxing. Following well established practice, we further reduce computing time by generating good starting points for the Branch \& Cut procedure.  Various approaches have been proposed to produce high quality warm starts (e.g., \cite{beck2013sparsity}, \cite{patrascu2015random}, \cite{hazimeh2018fast}) -- any of which could be used here -- but we opt for a simple and very inexpensive Forward Selection (FS) (\cite{hastie2005elements}). FS can be solved for exactly $k$ features, producing selection indicators $\+z_{FS}$ and coefficient estimates $\hat{\+\beta}_{FS}$. Running MIP with these as warm starts substantially cuts computational burden. 

While warm starts effectively reduce the MIP upper bound on the error, they do not affect the lower bound. When solving for a generic $k$, termination is triggered either by reaching a sufficiently small integrality gap between lower and upper bound (\cite{Schrijver1986}, \cite{Chen2011}), or by surpassing a maximum computing time (both rules can be customized in state-of-the-art solvers). In our implementation, as customary, we set a gap threshold $\epsilon_G$ as well as a $maxtime$. For the most expensive runs, $maxtime$ may be reached before the current gap is at or below $\epsilon_G$. Notably, the solver often reaches an optimal or near optimal point earlier than the total solution time may indicate (\cite{bertsimas2016best}, \cite{hastie2017extended}) due to a weak lower bound that is slow to improve. On the other hand, \cite{hastie2017extended} noted that the setting of a bound on computation time (3 minutes, in the specific case) may have contributed to the poorer performance of MIP compared to LASSO approaches in their more complex experiments. To mitigate these issues we resort again to FS, this time to create a surrogate lower bound. 

We rerun FS, now for $k+1$ features, and take the resulting in-sample MSE $\mathcal{E}_{FS^+}$ as an {\em error bound}. We continue solving our MIP and terminate when the surrogate gap between the current objective and $\mathcal{E}_{FS^+}$ is 
$\leq \epsilon_{FS^+}$ (surrogate gap threshold). This way, instead of relying on a potentially overly stringent $maxtime$ parameter (or setting an arbitrarily larger value to be more conservative), we continue the Branch \& Cut procedure until it generates a solution that is more parsimonious than that of FS (one fewer feature) and has comparable error (a gap matching $\epsilon_{FS^+}$). Figure~\ref{cvmseInncv}(b) shows 
a dramatic cut in computing time across folds compared to standard cross-validation. Here ``Integrated" refers to ICV alone, while ``Integrated+" and ``Integrated++" trigger the surrogate lower bound after a $maxtime$ of 3 and 0 minutes, respectively. Finally, to account for cases where the surrogate lower bound is still too hard to match within a realistic time frame, we set a more generous $totaltime$ parameter (Algorithm 2 in the Supplement provides pseudocode for ICV). 

\vspace{-0.2in}
\subsection{Whitening
}
\vspace{-0.1in}

With strong collinearities and/or weak signals, an accurate estimate of the size of the active set does not guarantee that solving for $M_{\hat{k}_o}$ will select the relevant features. 
Figure~\ref{whitening} and Figure~2 in the Supplement show results obtained solving with the true $k_0=10$ on data simulated for a range of Signal-to-Noise ratios (SNR; from $0.1$ to $10$) and highly collinear features. We considered an autoregressive correlation structure with $corr(\+X_\ell,\+X_j) = \alpha^{|\ell-j|}$, and a block structure with $corr(\+X_\ell,\+X_j)=\rho$ within the active and inactive sets, and $corr(\+X_\ell,\+X_j) =\omega$ across the two sets (see Section~4). For $\alpha=0.9$, or $\rho = 0.5$ and $\omega = 0.4$, MIP solutions lack sensitivity even with the true ${k}_0$, especially at low SNRs. However, we find that a {\em whitening} data pre-processing procedure 
can greatly improve MIP recovery.

In full generality, whitening involves switching from the original  $\*X$ ($n \times p$) to a new design matrix $\*Z = \*X \*W$ ($n \times p$), where the whitening matrix $\*W$ ($p \times p$) is chosen so that the transformed features are uncorrelated.
\begin{figure}[ht]
    \centering
        \begin{subfigure}[t]{0.49\textwidth}
              \includegraphics[scale=0.25]{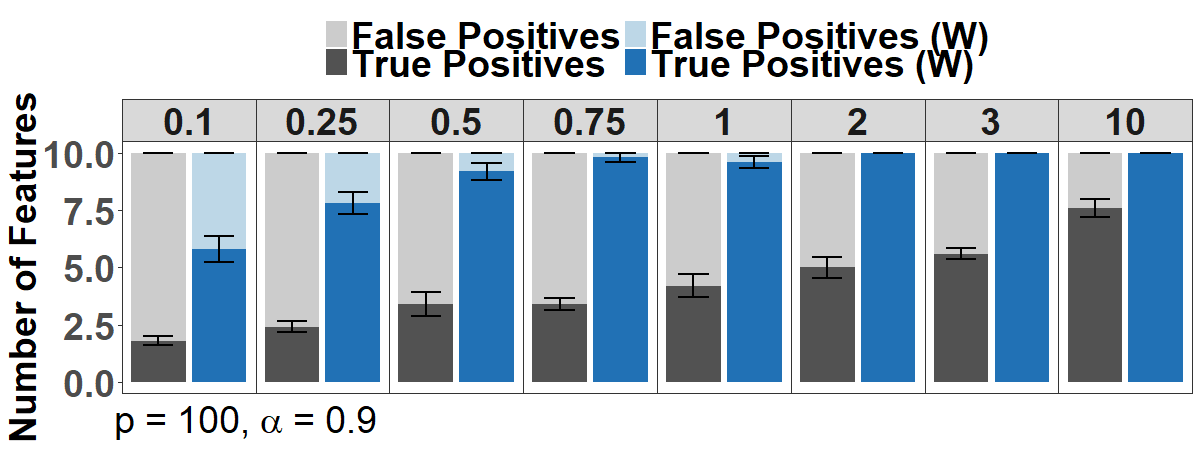}
              \caption{\textcolor{black}{Autoregressive correlation structure}}
          \end{subfigure}%
           ~
        \begin{subfigure}[t]{0.49\textwidth}
          \includegraphics[scale=0.25]{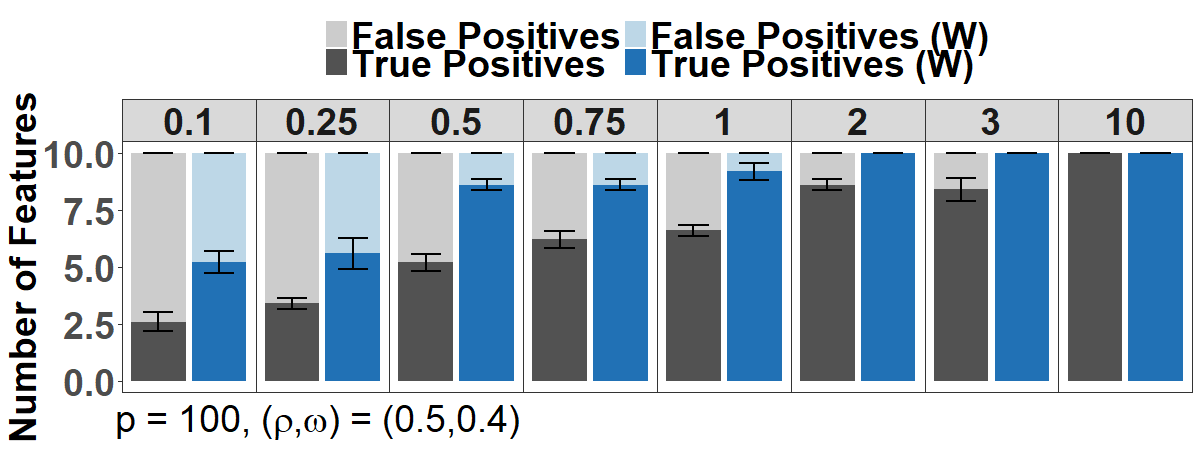}
          \caption{\textcolor{black}{Block correlation structure}}
        \end{subfigure}
        \vspace{-.3cm}
        \caption{Breakdown in average numbers of true and false positives selected solving MIPs across a range of SNR values (displayed along the top $x$-axis, error bars at $+/- 1SD$). Data are simulated from scenarios with $p=100$, $k_0=10$, and highly correlated features
        \color{black}
         under two correlation structures.}
         \color{black}
        \label{whitening}
\end{figure}
\vspace{-0.15in}

We use the sample covariance matrix and set $\*W=\hat{\*\Sigma}^{-1/2}$, which is known as ZCA or ZCA-Mahalanobis whitening. ZCA minimizes the distance between original and whitened features (\cite{kessy2018optimal}), and creates a direct feature pairing considering distances between the $j$th feature in $\*X$ and $\*Z$. MIP, or any other selection method, can then be applied to select whitened features and translate the selection back to the original features. The accuracy of this translation may vary based on the strength of the minimized distance but, as shown in Figure~\ref{whitening}, it reaches almost $100\%$ true positives at SNRs $\geq 0.75$ also under high collinearity. Post selection, one can use any estimator on the feature subset; in our experiments we simply use the OLS (Algorithm 3 in the Supplement provides pseudocode for whitening).

ZCA whitening has a marked positive effect also on the CVMSE curve (see Supplement; Section~3, Figure~3); it produces a more distinct elbow and solutions reasonably close to the true $k_0=10$. 
Interestingly, LASSO procedures do not seam to benefit from whitening as much as MIP, as they are prone to specificity rather than sensitivity issues (see Sections~2.3 and~4).
To implement whitening, we used the 
MLE of the covariance matrix in low-dimensional experiments (Figure~2 of the Supplement), where it 
is reasonably accurate, and the true covariance matrix in high-dimensional experiments. Investigating the use of effective high-dimensional covariance estimators in whitening (see \cite{fan2016overview} for an overview) exceeds the scope of this article and is left for future work.

\color{black}

\vspace{-0.2in}
\section{Numerical Studies}
\vspace{-0.1in}


To assess statistical quality, we compare MIP-BOOST to LASSO procedures and 
FS. To assess computational efficiency, we compare the burden of 10-fold cross-validation 
with MIP-BOOST with that of a standard MIP implementation (LASSO procedures are not considered here as they are orders of magnitude faster).

\vspace{-0.2in}
\subsection{Simulation Study}
\vspace{-0.1in}

We consider a variety of scenarios defined in terms of problem size $p$, sparsity level $k_0$ (the 
true
number of active features), signal patterns in $\+\beta$ (the regression coefficients), SNR levels, and nature and strength of feature collinearities.

We draw the $n$ rows of the data matrix $\*X$ independently from a $p$-variate Gaussian distribution $\*X \sim N_p(\*0,R)$ -- without loss of generality, means are set to $0$ and variances to $r_{j,j}=1$. For the correlations, we consider two regimes. The first is an {\em autoregressive structure} with $r_{\ell,j} = \alpha^{|\ell-j|}$ for $\ell \neq j$. Here each feature is strongly associated with a few neighboring ones -- and only weakly or almost not at all to features further away. If we list the active features as the first $k_0$, followed by the $p-k_0$ inactive ones, potential confounding across the former and the latter occurs for those ``within a certain radius" of the $k_0$ boundary -- depending on the size of $\alpha \in [0,1]$. The second regime is a {\em block structure} where all active features are pair-wise correlated at level 
$r_{\ell,j}=\rho$, $\ell, j \in A$, $\ell \neq j$, 
all inactive features are too, 
$r_{\ell,j}=\rho$, $\ell, j \in A^C$, $\ell \neq j$, 
and across the two sets one also has pair-wise correlations at level
$r_{\ell,j}=\omega$, $\ell \in A, j \in A^C$. 
Here the sizes of $\rho \in [0,1]$ and $\omega \in [0,1]$ control, respectively, the degree of potential confounding among active features, and between active and inactive ones. Notably, each active feature has the same strength of linear association with {\em all} inactive features; sizable values of $\omega$ induce strong violations of the irrepresentable condition (\cite{zhao2006model}), which in turn induce a loss of selection consistency for LASSO-type methods. These structures proxy different types of real data settings and have different eigenvalue decay profiles (see Figure~1 in the Supplement); with appropriately high parameter values, they can both represent strong collinearity.

The response vector is generated as $\*{Y} = \*X\+\beta + \+\epsilon$ where $\+\epsilon ~\sim~N_n(0,\theta^2I)$. Values of the scalar $\theta$ are used to create scenarios with varying SNRs, defined as SNR $= \frac{var(\+\beta^T \*X)}{\theta^2}$. A summary of all parameter settings is given in Table~\ref{parameters}. ``Realistic" SNRs may vary depending on scientific fields and class of problems. In recent literature, the MIP approach was shown to perform well at very high SNRs (\cite{bertsimas2017sparse}) and, conversely, to perform worse than LASSO procedures at very low SNRs (\cite{hastie2017extended}). Both extremes may characterize a limited number of real applications; we explore a range of SNR values from $0.1$ to $10$, corresponding to $R^2$ values from $\approx 1\%$ to $\approx 90\%$ (results below are for SNR $=0.1, 0.75, 1, 10$, full results are provided in the Supplement).
\begin{footnotesize}
\begin{table}[ht]
\footnotesize
\caption{Summary of parameter settings used in the simulation study}\label{parameters}
\begin{threeparttable}
\begin{tabular}{lll}
\hline
Parameter & Values \\ \hline
$n\ :\ p\ :\ k_0$ & $500\ :\ 100,1000\ :\ 10$ \\ 
$\alpha$, AUTOREG & *0.8, 0.9 \\ 
$(\rho,\omega)$, BLOCK & *(0.5,0.1), *(0.5,0.3), *(0.5,0.4) \\ 
$\+\beta$ & 
$(1,1,1,1,1,1,1,1,1,1,0,...,0)^T$, **$(10,10,10,10,10,10,10,5,5,5,0,...,0)^T$ \\ 
SNR ($R^2$) &  0.1 (0.09), *0.25 (0.20), *0.50 (0.33), 0.75 (0.43), \textbf{1 (0.50)}, *2 (0.67), *3 (0.75), 10 (0.91) \\
\hline  
\end{tabular}
\begin{tablenotes}
\item *: results in Supplement. **: 7 stronger and 3 weaker active signals ($\beta_j=10$ and $5$). Scenarios are replicated 5 or 10 times depending on the SNR. $R^2$ based on the SNR using the formulation in \cite{hastie2017extended}.
\end{tablenotes}
\end{threeparttable}
\end{table}
\end{footnotesize}

\vspace{-0.2in}
\subsubsection{Statistical Quality}
\vspace{-0.1in}
Fixing the number of cross-validation folds to $10$, we compare our MIP-BOOST against the {\em standard LASSO} with the minimizing or the parsimonious choice of $\lambda$ (LMN and LSD, see Section~2.3); the {\em relaxed LASSO} of \cite{hastie2017extended} with the minimizing choice of $\lambda$ (RL); and {\em standard forward selection} with the minimizing choice of subset size (FS). For LMN, LSD, RL and FS we use the \texttt{R} package \texttt{bestsubset} from \cite{hastie2017extended}. 
For MIP-BOOST we use \texttt{Julia 0.6.1}
to interface with the MIP optimization solver \texttt{Gurobi 7.5.1}. 
$maxtime$ is set to the recommended $3$ minutes for all scenarios, but $totaltime$ is set to $10$ and $20$ minutes for scenarios with $p=100$ and $p=1000$, respectively
\footnote{Calculations were performed on the Institute for CyberScience Advanced CyberInfrastructure (ICS-ACI) of Penn State, using the basic memory option on the ACI-B cluster with an Intel Xeon 24 core processor at 2.2 GHz and 128 GB of RAM. The multi-thread option in \texttt{Gurobi} was limited to a maximum of 5 threads for consistency across settings.}. 
Optimality and surrogate gaps are set to $\epsilon_G=0.05$ and $\epsilon_{FS+} = 0.05$ (see Section~\ref{BWF}). 
\color{black}
Concerning the ``big M" parameters (see Section~2.3), we set $\mathcal{M}_j = c|\hat{\beta}_{j,OLS}|$ where $\hat{\+\beta}_{OLS}$ is the vector of Ordinary Least Squares (OLS) estimates and we set the scaling constant to $c=5$ 
(inflating by $c$ does not give rise to solvability issues).
\color{black}
MIP-BOOST (\texttt{Julia}) and data generation (\texttt{R}) code is available upon request 
\footnote{A package for general use is forthcoming.}.

Simulation scenarios are replicated $5$ or $10$ independent times. As performance metrics, we consider (i) average number of true and false positives ({\em sensitivity} and lack of {\em specificity}); (ii) average MSE computed on separately drawn validation data for each scenario and simulation ({\em prediction accuracy}); and (iii) accuracy in estimating $\+\beta$, partitioned into {\em variance} 
and {\em squared bias}. Results for (i) and (ii) are shown below and those for (iii) in Section~4 of the Supplement. We report the metrics with/without whitening depending on whether it helps or hurts each competing approach; in the scenarios considered LMN, LSD and RL perform better without whitening, and results for FS vary depending on SNR, covariance structure and problem size.

\begin{figure}[ht]
	\centering
	\begin{tabular}[c]{cc}
         \begin{subfigure}[c]{0.5\textwidth}
              \includegraphics[scale=0.23]{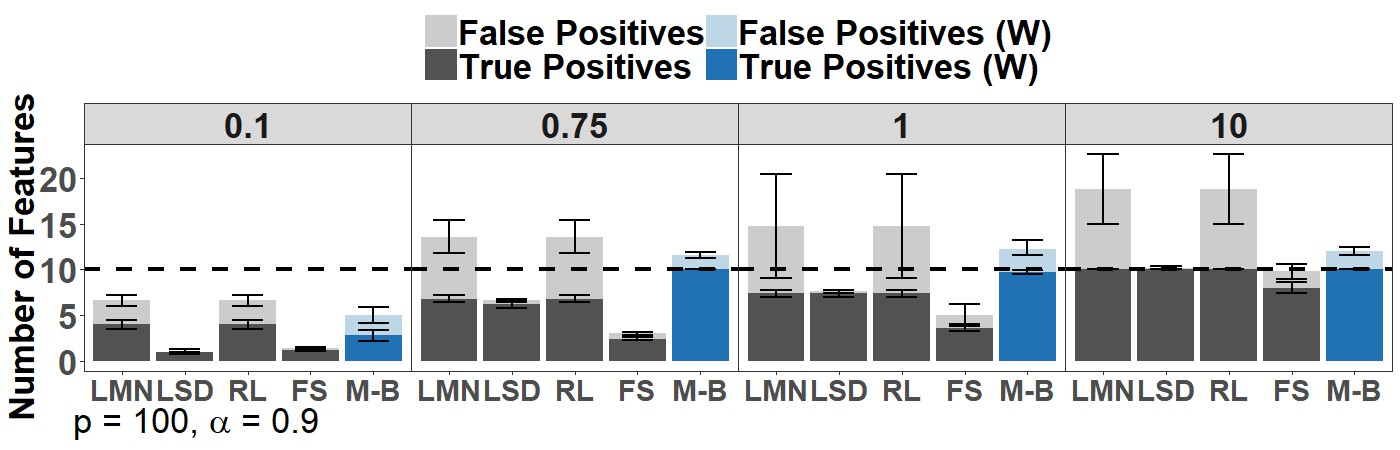}
              \caption{
              Average number of true and false positives over simulation replicates. Bars: $\pm 1SD$, dashed line: $k_0=10$. 
              }\label{TFP-LD-09-COMBO} 
          \end{subfigure} &
              \begin{subfigure}[c]{0.5\textwidth}
              \includegraphics[scale=0.23]{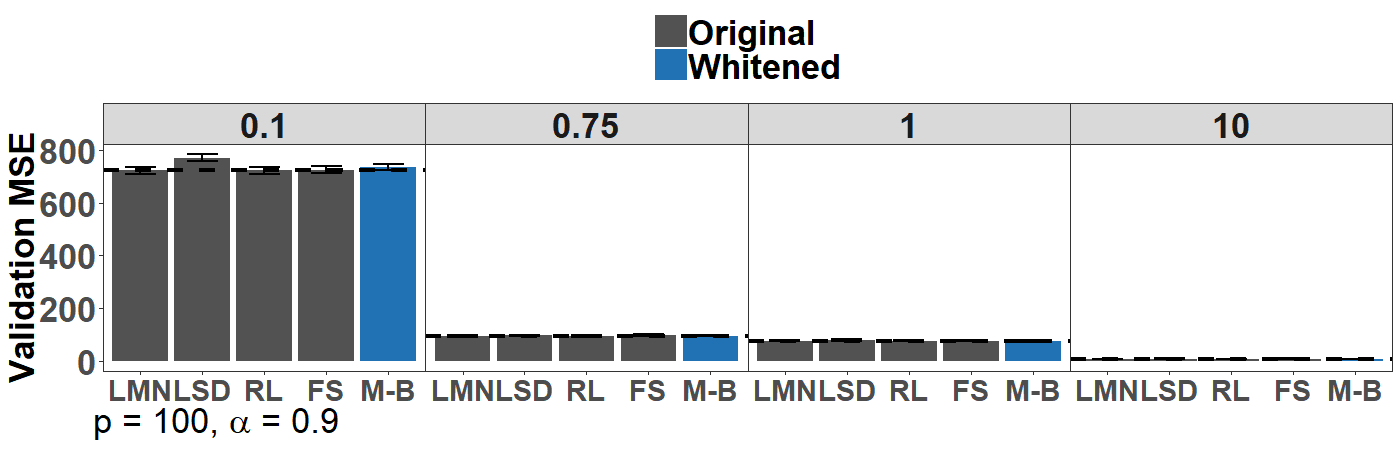}
              \caption{
              Average validation MSE over simulation replicates (error bars at $+/- 1SD$). Dashed line: error with an OLS fit on the relevant features.
              }\label{PMSE-LD-09-COMBO}
          \end{subfigure} \\
         \begin{subfigure}[c]{0.5\textwidth}
              \includegraphics[scale=0.23]{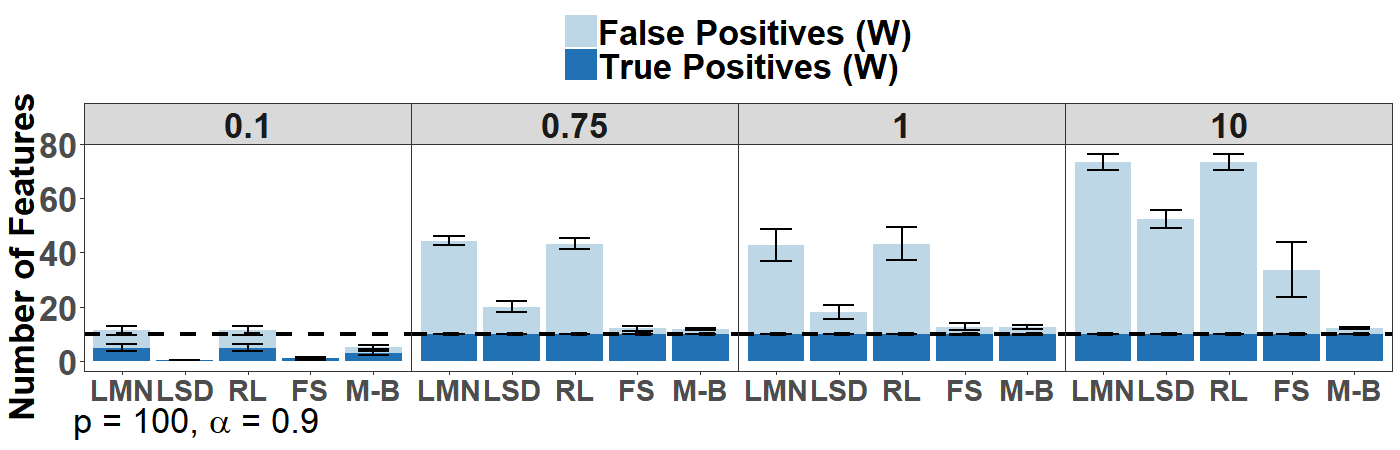}
              \caption{Average number of true and false positives over simulation replicates. Bars: $\pm 1SD$, dashed line: $k_0=10$.}
              \label{TFP-LD-09-WHITE-01}
        \end{subfigure} &
         \begin{subfigure}[c]{0.5\textwidth}
            \includegraphics[scale=0.23]{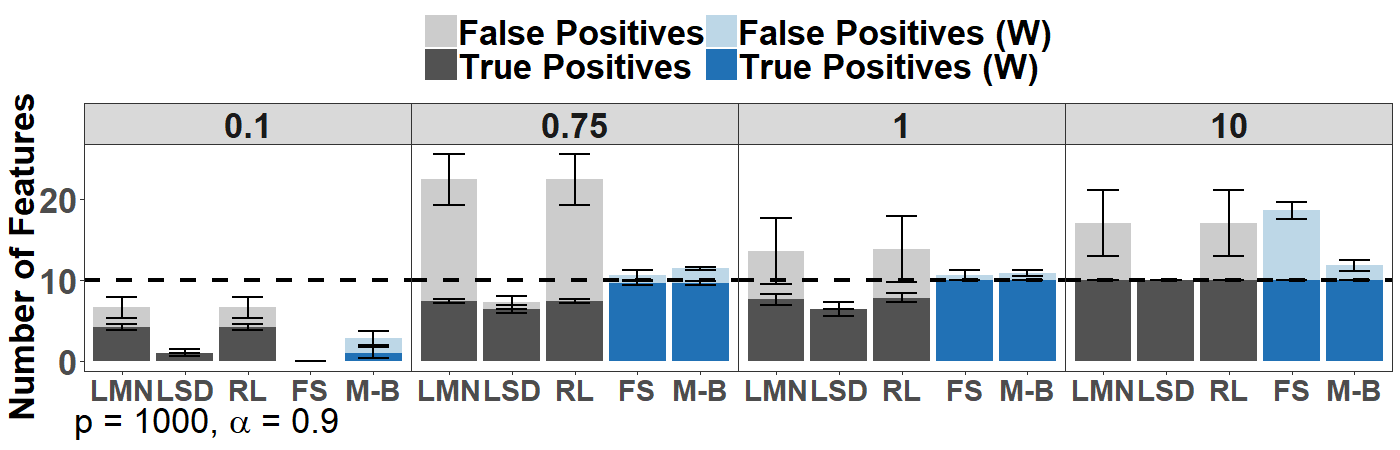}
            \caption{
            Average number of true and false positives over simulation replicates. Bars: $\pm 1SD$, dashed line: $k_0=10$.}\label{TFP-HD-09-COMBOTW}
        \end{subfigure} \\
          \begin{subfigure}[c]{0.5\textwidth}
            \includegraphics[scale=0.23]{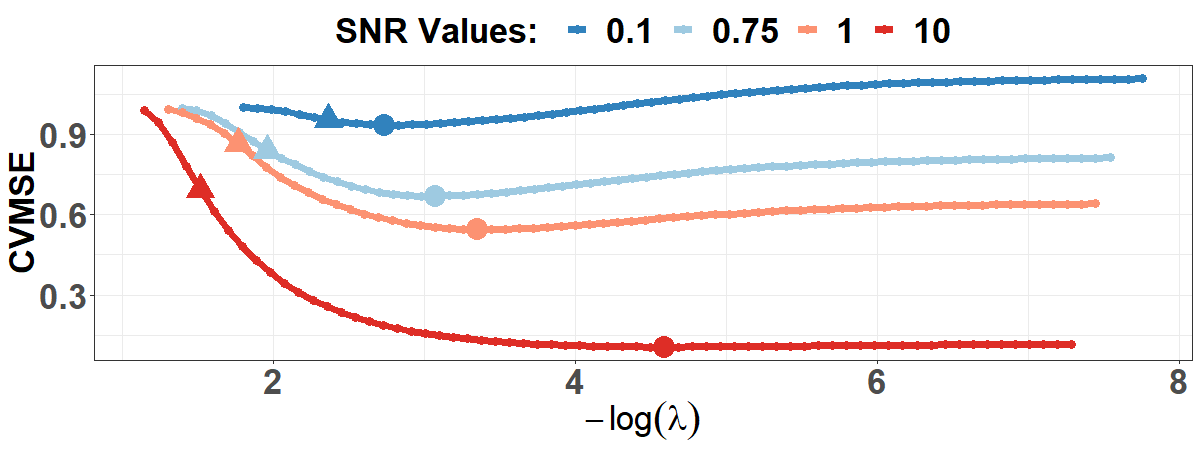}
            \caption{LASSO cross-validation. $\bigtriangleup:\lambda$'s for subsets of size closest to $k_0=10$. $\bigcirc:\lambda$'s minimizing CVMSE.}
            \label{LASSO-CV-SNR-LD-0504}
        \end{subfigure}&
        \begin{subfigure}[c]{0.5\textwidth}
            \includegraphics[scale=0.23]{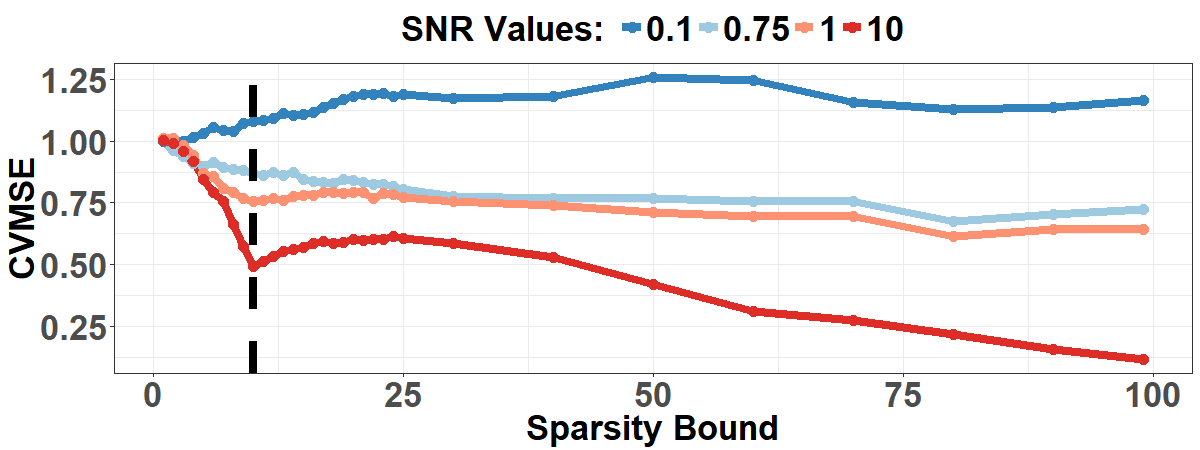}
        \caption{MIP-BOOST cross-validation. 
      Dashed line: $k_0=10$.}\label{MIQP-CV-SNR-LD-0504}
      \end{subfigure}\\
	\end{tabular}
    \caption{Summary results in scenarios with $n=500$, $p=100$ (panels (a)-(c)), $p=1000$ (panel (d)), $k_0=10$, various SNR values, and autoregressive correlation structure with $\alpha=0.9$. M-B stands for MIP-BOOST in the bar charts. Gray: original data. Blue: withened data (MIP-BOOST for $p=100$ in (a) and (b), all procedures in (c), MIP-BOOST and FS in (d)).
    We used 10 replicates for SNR $\leq 0.75$ and $p=100$, and 5 for all other scenarios. Cross-validation behavior: LASSO in (e) and MIP-BOOST in (f).
}
\label{LDHD-09}
\end{figure}

\vspace{-0.2in}
\paragraph{Scenarios with $p=100$:}
Here the dimension $p=100$ is one fifth of the sample size $n=500$, and the number of active features is $k_0=10$. Figure~\ref{LDHD-09} summarizes results with autoregressive correlations ($\alpha=0.9$) and SNRs around 0.1, 0.75, 1 and 10 (plots for all SNRs and both correlation structures are shown in the Supplement). 

Figure~\ref{LDHD-09}(a) shows that all procedures capture a limited number of true positives when the SNR is low. 
LMN and RL select a few more true positives than other procedures, but are still both too sparse and unspecific.
The lack of specificity of LMN and RL does not improve (it worsens) at higher SNRs. 
In contrast, the parsimonious LSD is rather specific -- but it, too, 
struggles to capture true positives unless signals are strong. MIP-BOOST shows some lack of specificity, but in comparison to LMN and RL this is milder and less dependent on signal strength. In addition, it already captures all true positives at an SNR as small as $0.5$ (Figure~5, Supplement). FS, while appearing rather specific like LSD, catches up more slowly than all others in capturing true positives. Given the strong collinearities, these differences in sensitivity and specificity do not translate in substantial differences in prediction accuracy, as shown in Figure~\ref{LDHD-09}(b).


Figure~\ref{LDHD-09}(c) shows true and false positive results when all procedures are applied to whitened data. 
Comparing this to Figure~\ref{LDHD-09}(a), the already unspecific LASSO procedures become even more unspecific. FS, too sparse but very specific on the original data, becomes even sparser at low SNRs and loses its specificity at high SNRs. This loss of specificity is yet more pronounced under the block correlation structure (Figure~3 in the Supplement; FS selects all $100$ features at SNR $\geq 2$). 

Figures~\ref{LDHD-09}(e) and \ref{LDHD-09}(f) provide insight on why whitening affects LASSO and MIP-BOOST 
differently. With whitening the LASSO CVMSE curves give little information on the location of the optimal $\lambda$. As the SNR increases, the $\lambda$ minimizing the error gets further and further away from the $\lambda$ inducing the right level of sparsity (i.e.~corresponding to $k \approx 10$). At SNR$=0.1$ the two $\lambda$'s are close, but this still corresponds to a solution where more than half of the selected features are false positives (Figure~\ref{LDHD-09}(c)). 
In contrast, the MIP-BOOST CVMSE curves point more clearly to the right level of sparsity as the SNR increases. Curves shown here 
are generated by solving on $10$ folds across an entire grid of $k$ values ranging from $1$ to $100$
\footnote{
For SNR $10$, MIP-BOOST was run with a more conservative threshold set for BF; a quick run of the LASSO selecting a very large number of features can inexpensively diagnose the need to implement a restrictive BF. However, we stress that BF used in MIP-BOOST does {\em not} suffer from the same lack of specificity that hinders LASSO at high SNRs. When the SNR increases the true $k_0$ becomes in fact easier to identify for MIP; it is simply that the BF search may be pushed towards the right of the $k$ range if seemingly large differences in CVMSE are not ''ignored'' by setting a stricter threshold. In Figure~\ref{LDHD-09}(e), the elbow at $10$ is clearly denoted when SNR$=10$, but the strong signal leads to distinct drops in the CVMSE also at large $k$ values. A more restrictive threshold easily improves solution quality. In contrast, using a more conservative tuning for LASSO (e.g.,~the parsimonious LSD), still fails to increase specificity.}.

\vspace{-0.2in}
\paragraph{Scenarios with $p=1000$:} the dimension $p=1000$ is twice the sample size $n=500$ (undersampling). The number of active features remains $k_0=10$. Figure~\ref{LDHD-09}(d) summarizes true and false positive results with autoregressive correlations ($\alpha = 0.9$) and SNRs $0.1, 0.75, 1$ and $10$ (complete results are again shown in the Supplement). For LASSO procedures and MIP-BOOST, results are remarkably and reassuringly similar to those for $p=100$.
However, the behavior of FS (which here performs best on whitened data) changes notably with respect to scenarios with $p=100$. 

\vspace{-0.2in}
\paragraph{Other scenarios (see Supplement):}
We end noting that results with the block structure are consistent with those with the autoregressive structure.
However, as the correlation between active and inactive features increases all methods become less effective, even under whitening -- likely due to violations of irrepresentability (\cite{zhao2006model}).
To test a lower dimensional but sparser scenario we also simulated data with the block structure for $p=100$ and $k_0=2$. MIP-BOOST is still more specific than LMS and RL and, surprisingly, only LSD benefits from increased sparsity. 

\vspace{-0.2in}
\subsubsection{Computational Burden}
\vspace{-0.1in}
Next, we document 
the dramatic cut in computing time afforded by MIP-BOOST vs.~a standard, or naive, MIP implementation -- which selects $\hat{k}_0$ searching the entire sparsity bound range with traditional cross-validation. This cut renders a rigorously tuned MIP viable
in a broader variety of realistic scenarios. We measure the total time required for a 10-fold cross-validation search of the interval between $1$ and a maximum sparsity bound $c$, comparing naive MIP, MIP-BOOST with no surrogate lower bound stopping (Section~3.2), MIP-BOOST+ where the surrogate lower bound is triggered after reaching a 3 minutes $maxtime$, and MIP-BOOST++ where it is applied it immediately. We fix again $totaltime=10$ and $20$ minutes for $p=100$ and $p=1000$, respectively, and allow 1 thread per instance to mimic settings in previous results (\cite{hastie2017extended}).

A conservative estimate of the speedup expected from MIP-BOOST if the computing time required for each MIP evaluation were equivalent to that of naive MIP and equal across the entire search interval is obtained comparing the total number of sparsity bounds explored in traditional cross-validation (say, $c$) and in the proposed bisection (at most $log_2(c)+3$) -- see Supplement, Section~5, Figure~26.
The benefits of MIP-BOOST become more dramatic the larger is $c$, which is important because the sparsity bound range to be explored increases with the dimension $p$ of the problem. Moreover, actual MIP-BOOST speedups (see below) vastly surpass this conservative estimate.

\begin{figure}[H]
	\centering
	\begin{tabular}[c]{cc}
        \begin{subfigure}[c]{0.5\textwidth}
              \includegraphics[scale=0.23]{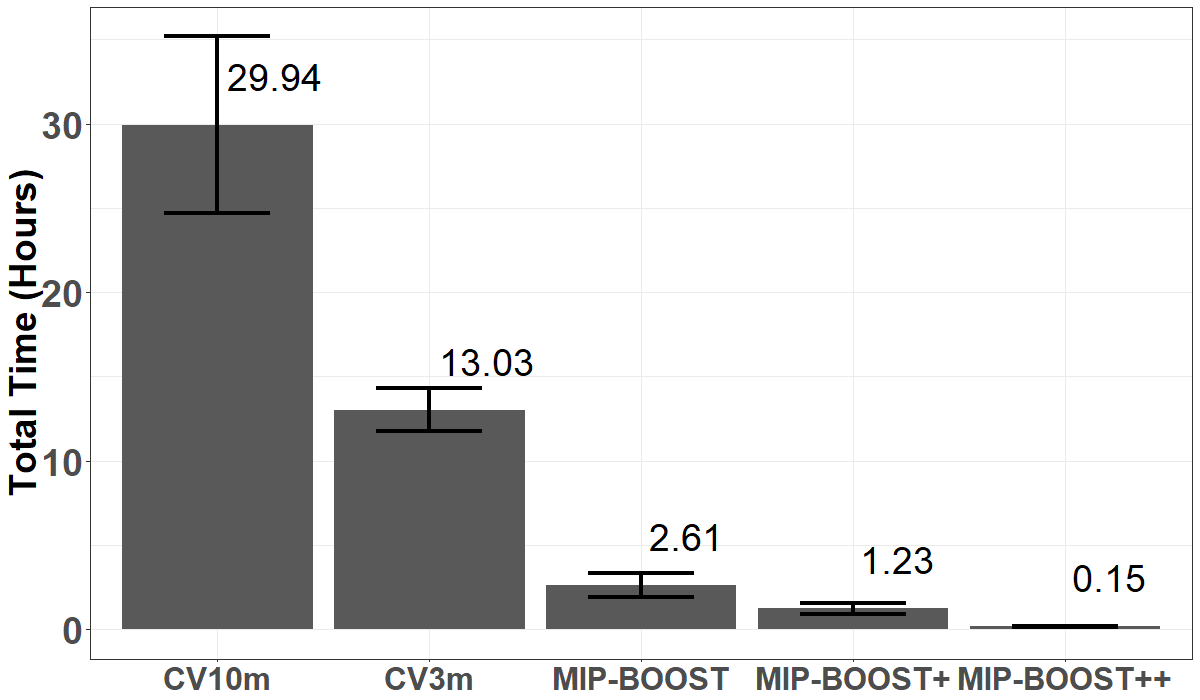}
              \caption{
              Average total time (in hours). Bars: $\pm 1SD$. 
              }\label{TT-LD}
          \end{subfigure}&
        \begin{subfigure}[c]{0.5\textwidth}
            \includegraphics[scale=0.23]{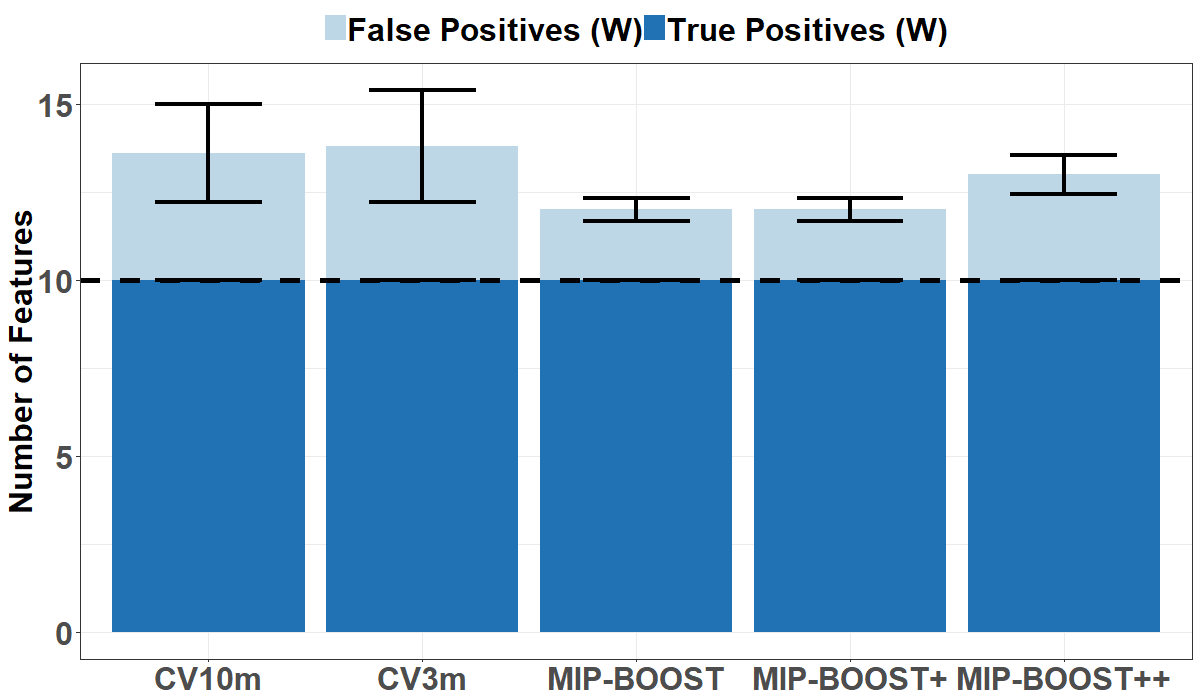}
            \caption{Average no. true and false positives. Bars: $\pm 1SD$, dashed: $k_0=10$.}
            \label{TFP-LD}
        \end{subfigure}\\
         \begin{subfigure}[c]{0.5\textwidth}
              \includegraphics[scale=0.23]{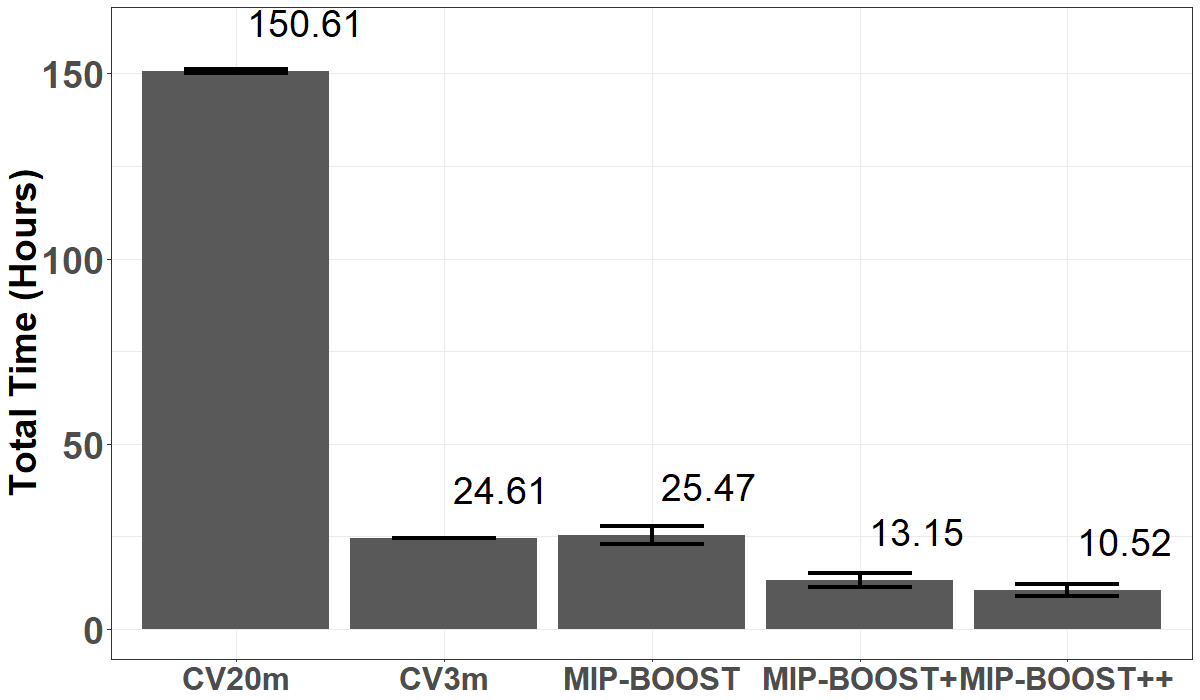}
              \caption{Average total time (in hours). Bars: $\pm 1SD$.}
              \label{TT-HD}
        \end{subfigure}&
      \begin{subfigure}[c]{0.5\textwidth}
            \includegraphics[scale=0.23]{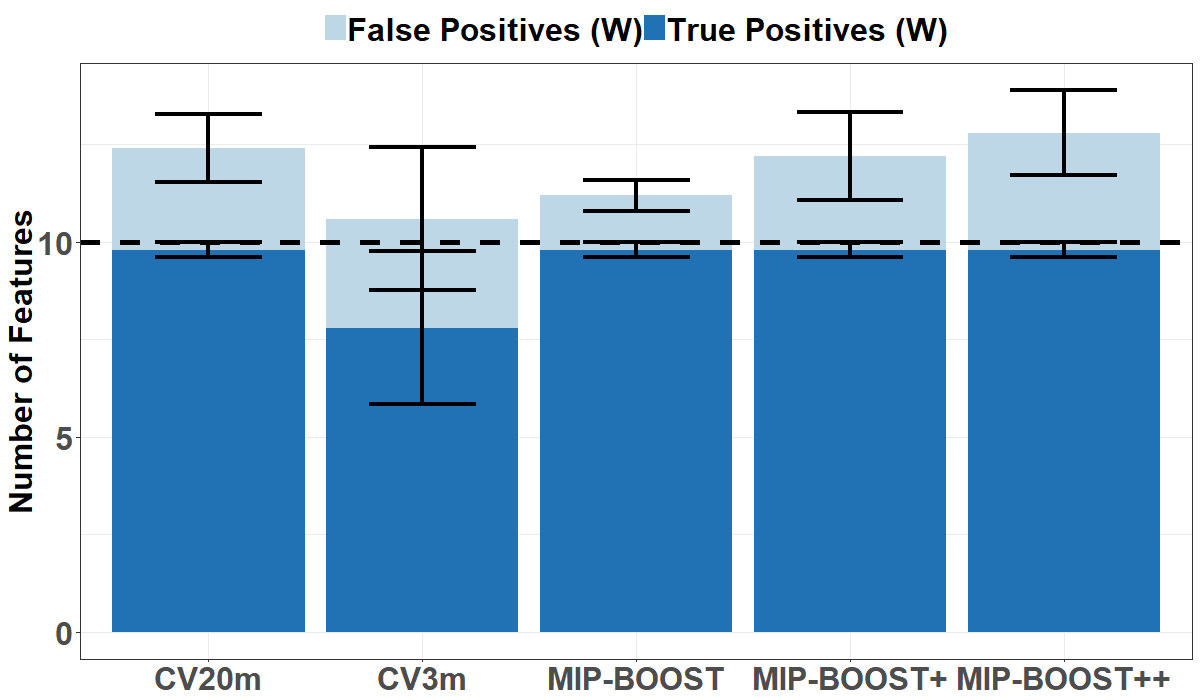}
            \caption{Average no. true and false positives. Bars: $\pm 1SD$, dashed: $k_0=10$.}
            \label{TFP-HD}
        \end{subfigure}\\
    \end{tabular}
    \caption{Summary results in scenarios with $n=500$, $p=100$ (panels (a)-(b)); $p=1000$ (panels (c)-(d)); $k_0=10$, autoregressive correlations ($\alpha=0.9$) and SNR $=1$. We used 5 replicates.
    }
\label{TT-09}
\end{figure}

Figures~\ref{TT-09}(a)- \ref{TT-09}(c) show results for a 
scenario with $n=500$, $p = 100$, $k_0=10$, autoregressive correlations ($\alpha=0.9$), and SNR $=1$. The range considered extends to $c=100$. The naive MIP uses traditional 10-fold cross-validation, with a maximum computing time per MIP evaluation set to 10 (CV10m) or 3 (CV3m) minutes. 
On average, MIP-BOOST is 11.5x faster then CV10m and 5x faster than CV3m. MIP-BOOST+ and MIP-BOOST++ are 24.3x and 199.6x faster than CV10m, respectively. Importantly, the best upper bound (in-sample error objective) found per instance is essentially the same across procedures (see Supplement, Section~5, Figure~27(a)). 
In fact, if we omit the bisection and solve across all 100 sparsity bounds, the average upper bound of MIP-BOOST is equal or stronger than that of CV10m and CV3m for 75\% and 81\% of the values between 1 and $c=100$, respectively (64\% and 77\% for MIP-BOOST+, 28\% in both cases for MIP-BOOST++). 
Across the sparsity bound range, the drop in quality of MIP-BOOST++ averages $\sim 0.5\%$ and is never higher than $\sim 2\%$ (see Supplement, Section~5, Table~1 and Figure~28(a)). Additionally, Figure~\ref{TT-09}(b) shows that the cut in computational burden does {\em not} come at the expense of statistical quality; in fact, all MIP-BOOST variants favor sparser solutions with fewer false positives than CV10m and CV3m. 

\vspace{-.15in}
\vspace{-0.2in}

In Figures~\ref{TT-09}(c) and \ref{TT-09}(d) we raise the dimension to $p=1000$, leaving all other simulation parameters untouched. We limit the range considered to $c=50$ to keep traditional cross-validation viable, and increase the maximum computing time per MIP evaluation to 20 minutes (CV20). 
Here MIP-BOOST variants are as much as 14.3x faster than CV20m.
MIP-BOOST+ and MIP-BOOST++ are also distinctly faster. 
The average upper bound of all three MIP-BOOST variants is equal or stronger than that of CV20m and CV3m for 72-96\% and 98\% of the sparsity bounds, respectively (see Supplement, Section~5, Table~1 and Figure~28(b)). For larger sparsity bounds, CV3m often fails to find an integer solution within the time limit; the missing CVMSE values make CV3m difficult to tune. 
We can still tally true and false positives based on the sparsity bound with minimum CVMSE among those that produced solutions in each fold, but remembering that the search options are biased towards smaller values. 
With this caveat, Figure~\ref{TT-09}(d) shows the poor statistical quality of CV3m, with a large and highly variable number of false positives even when choosing a bound close to the true $k_0=10$. Here MIP-BOOST is the most effective, followed by MIP-BOOST+;
MIP-BOOST++ selects marginally more false positives than CV20m while being 14.3x faster. Thus MIP-BOOST is recommended for accuracy, even in comparison to CV20m. MIP-BOOST+/++ are yet cheaper -- but at the expense of 
a little reduction in statistical quality.


The results above compare computing time and upper bounds under a restricted maximum computing time per MIP evaluation (20, 10, or 3 minutes). With $p=1000$, this maximum was reached in the majority of cases and by all procedures. To further demonstrate the effectiveness of MIP-BOOST vs.~naive MIP, we compared the time to reach an optimality gap of 5\% between traditional cross-validation and ICV at the true $k_0=10$, across the 10 cross-validation folds. 
Here the SNR was raised to $10$. This was repeated on 10 different simulated data sets, totaling 100 MIP runs for the 2 procedures. ICV converged faster in 75 of the 100 cases - 
\color{black}
providing a median time reduction of 13\%.
\color{black}
\vspace{-0.2in}
\subsection{An Application to Diabetes Data}
\vspace{-0.1in}
We analyze the diabetes dataset first used in \cite{efron2004least}. This is a rather low dimensional and well sampled dataset, containing $n=442$ diabetes patients and $10$ baseline predictors (age, body mass index, average blood pressure, six blood serum measurements and sex encoded as $\{0,1\}$). 
The response is a measure of disease progression one year after the baseline. The number of features considered is actually $p=64$; in addition to the $10$ predictors, these include all their $45$ pair-wise interactions (products) and $9$ squared terms (the squared term for sex is not included). 
After splitting the data into training and test sets (354 or 75\%, and 88 or 25\% of the observations), we ran all procedures on both whitened and standardized training data. The latter are transformed to have mean $0$ and variance $1$ without eliminating correlations (both LASSO and MIP-BOOST must use features with equal scaling; \cite{friedman2001elements}). On the test set, we computed validation MSEs, again after whitening or simple standardization.
Results are provided in Table~\ref{DiabetesResults}, which shows also OLS fits of the null and full models (intercept only, and intercept plus all $64$ features). The last column contains the relative decrease in validation MSE between the full model and the models produced by each of the procedures considered. 
\begin{table}[h]
\footnotesize
\caption{Summary results for procedures applied to the Diabetes data set.
}\label{DiabetesResults}
\begin{tabular}{l|cc|cc|cc}
\hline
\textbf{Procedure} & \multicolumn{2}{c|}{\textbf{$\hat{k}_{0}$}} & \multicolumn{2}{c|}{\textbf{ValMSE}} &  \multicolumn{2}{c}{\textbf{Relative Decrease vs Full}}\\
 & \multicolumn{2}{c|}{\textbf{(S and W)}} & \multicolumn{2}{c|}{\textbf{(S and W)}} &  \multicolumn{2}{c}{\textbf{(S and W)}}\\ \hline
LMN & 19 & 41 & 0.488 & 0.631 & 0.614& 0.501\\
LSD & 9 & 21 & 0.501 & 0.550& 0.604& 0.565\\
RL & 19 & 41 & 0.488 & 0.631 & 0.614 & 0.501\\
FS & 16 & 7 & 0.523 & \textbf{*0.477} & 0.586 & 0.623\\
MIP-BOOST & 2 & 3 & 0.513 & 0.496 & 0.594 & 0.608\\
OLS Null & \multicolumn{2}{c|}{0} & \multicolumn{2}
{c|}{0.989} & \multicolumn{2}{c}{--}  \\
OLS Full & \multicolumn{2}{c|}{64} & \multicolumn{2}{c|}{1.265}
& \multicolumn{2}{c}{--} \\
\hline  
\end{tabular}
\begin{tablenotes}
\item *: minimum Validation MSE. 
\textbf{S}: standardization. \textbf{W}: whitening.
\end{tablenotes}
\end{table}

MIP-BOOST generates a very sparse solution, whose prediction accuracy does improve with whitening.
Whitening adds one feature, Lamotrigine (LTG), to the two already selected with standardization, Body Mass Index (BMI) and Mean Arterial Pressure (MAP). These all have highly significant p-values
(BMI $9.25 \times 10^{-14}$, LTG $2.67 \times 10^{-13}$ and MAP $2.05 \times 10^{-4}$). 
FS with whitening produces the minimum validation MSE, but with as many as $7$ features -- the three identified by MIP-BOOST, plus HDL, TCH, GLU and the interaction between BMI and MAP. This induces a modest improvement; in terms of relative decrease in the last column, only 1.5\%. p-values for BMI, LTG, MAP and the interaction between BMI and MAP are highly significant. In contrast, p-values for TCH and GLU, albeit selected, are non-significant (0.742 and 0.458, respectively) and HDL is only marginally significant (0.0302).
LASSO-type methods select many more features, but again with modest improvements in variance explained in the training set, and with a substantial deterioration in validation MSE. Notably, LMN and the RL have the same performance; relaxing the LASSO (\cite{hastie2017extended}) does not improve it on this data.
\vspace{-0.1in}
\begin{figure}[H]
	\centering
	\includegraphics[scale=0.33,trim={0 0 0 1cm},clip]{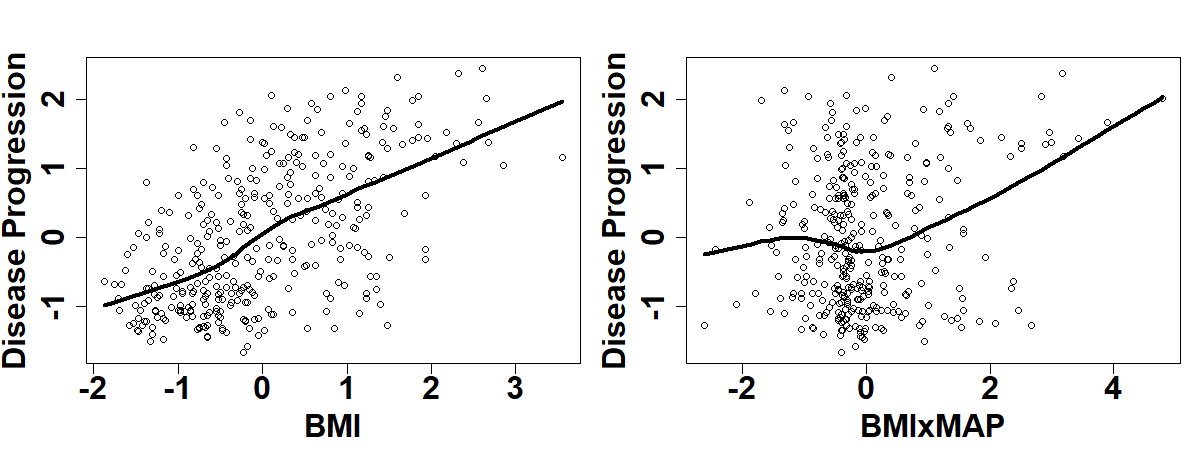}
    \caption{Scatter plots (with loess curves) 
    for Disease Progression against BMI (left) and BMIxMAP (right). The interaction shows a 
    less clear-cut effect, which may unfold mostly at very large values.
    }\label{Interaction}
\end{figure}
\vspace{-0.3in}
BMI, MAP and LTG are selected by all procedures and have very low p-values in all model fits. They were also included in the best models of \cite{efron2004least}. The interaction BMI$\times$MAP is not selected by MIP-BOOST, but it is selected by all other procedures and significant in the model fits. Its p-value in a $4$ feature model comprising BMI, MAP, LTG and BMI$\times$MAP is also quite low ($\sim 0.003$). So why did MIP-BOOST not select this term? The right panel of Figure~\ref{Interaction} shows a discernible marginal effect on disease progression, especially at very high values of BMI$\times$MAP. But the improvement in terms of validation MSE is negligible (relative decrease of 0.495 vs 0.496), as is the one in terms of variance explained in the training set ($R^2$ of 0.484 vs 0.472). For comparison, the left panel of Figure~\ref{Interaction} shows the much stronger and clearer-cut marginal effect of BMI on disease progression. Of course on a real data problem we have no conclusive way to confirm that we have captured all active features, but we have shown that MIP-BOOST can produce sparse, interpretable and competitive solutions -- and do so in a reasonable amount of time (with the relatively small $p=64$, MIP-BOOST total computing time was $\sim 13$ and $\sim 25$ minutes for whitened and standardized data, respectively).

\vspace{-0.2in}
\section{Concluding Remarks and Future Work}
\vspace{-0.1in}

Mixed Integer Optimization for 
$L_0$ feature selection is receiving much attention due to its foundational appeal and versatility. The effectiveness of this approach, like that of others, depends critically on the selection of a tuning parameter. 
Pursuing this through cross-validation requires the solution of a large number of MIP instances. Notwithstanding recent improvements in algorithmic efficiency, this can hinder computational viability in problems of realistic size and complexity. MIP-BOOST cuts computational burden by several orders of magnitude, with increasing gains as the dimension of the  problem, and thus the tuning parameter range to explore, increase. 

Importantly, though our proposals were described for linear models, they could easily be extended to generalized linear models and other classes of statistical models. Also importantly, with MIP as well as other methods, 
strong feature collinearity and/or weak regression signals can deteriorate 
statistical quality even when the optimization is well tuned. 
MIP-BOOST addresses collinearity through ZCA whitening.
This is not a new concept, but to our knowledge we are the first to demonstrate empirically its impact on feature selection.
Whitening is attracting increasing attention; \cite{kessy2018optimal} show how different forms of whitening can benefit different statistical problems. 
While our numerical studies show that whitening improves recovery in MIP procedures, they do not show a positive effect on LASSO performance. This may be due to the fact that LASSO suffers from poor specificity, not poor sensitivity. 
The effects of whitening on features selection methods deserve further investigation. In particular, with regards to the MIP approach, it will be critical to investigate whether and how fast an increase in the distance between original and ZCA whitened features erodes the gains in recovery. We also noted how covariance estimation accuracy can affect the effectiveness of whitening, something that is of special concern in high-dimensional settings. In this respect, we plan to explore different covariance and precision estimation approaches ( \cite{fan2016overview}). 

We envision several other avenues for future work. In our simulations, strong signals induced false positives for MIP-BOOST which were easily prevented with stricter improvement thresholds in the bisection -- suggesting that this may benefit from data-driven adaptive thresholding. Moreover, we could increase efficiency incorporating new techniques to generate cuts within the Branch \& Bound algorithm. \cite{bertsimas2017sparse} proposed a cutting plane method for sparse ridge regression which, under mild feature collinearity and fairly strong signals, showed good feature selection performance and low computational burden. Also, \cite{bertsimas2017logistic} proposed adding constraints to balance specific, competing goals of the modeler. Along similar lines, structure in the data could be used to improve solution quality and reduce computation.
\vspace{-0.3in}
\bibliographystyle{JASA}
\bibliography{Bibliography-MM-MC}
\end{document}